\documentclass[a4paper,11pt]{article}
\usepackage{pos}
\usepackage{xspace}
\usepackage{csquotes}
\newcommand{\ppbar}        {\mbox{\ensuremath{\mathrm {p\overline{p}}}}\xspace}

\newcommand{\ee}           {\ensuremath{\mathrm{e^{+}e^{-}}}\xspace}

\newcommand{\snn}          {\ensuremath{\sqrt{s_{\mathrm{NN}}}}\xspace}
\newcommand{\pt}           {\ensuremath{p_{\rm T}}\xspace}

\newcommand{\RpPb}         {\ensuremath{R_{\rm pPb}}\xspace}

\newcommand{\Raa}          {\ensuremath{R_\mathrm{AA}}\xspace}

\newcommand{\zjet}{\ensuremath{z_{\mathrm{||}}^{\mathrm{ch}}}\xspace}
\newcommand{\nineH}        {$\sqrt{s}~=~0.9$~Te\kern-.1emV\xspace}
\newcommand{\seven}        {$\sqrt{s}~=~7$~Te\kern-.1emV\xspace}
\newcommand{\twoH}         {$\sqrt{s}~=~0.2$~Te\kern-.1emV\xspace}
\newcommand{\twosevensix}  {$\sqrt{s}~=~2.76$~Te\kern-.1emV\xspace}
\newcommand{\five}         {$\sqrt{s}~=~5.02$~Te\kern-.1emV\xspace}
\newcommand{\twosevensixnn}{$\sqrt{s_{\mathrm{NN}}}~=~2.76$~Te\kern-.1emV\xspace}
\newcommand{\fivenn}       {$\sqrt{s_{\mathrm{NN}}}~=~5.02$~Te\kern-.1emV\xspace}

\newcommand{\GeVc}         {\ensuremath{\mathrm{GeV}/c}\xspace}

\newcommand{\TeV}          {\ensuremath{\mathrm{TeV}}\xspace}
\newcommand{\GeV}          {\ensuremath{\mathrm{GeV}}\xspace}

\newcommand{\vtwo}          {\ensuremath{v_\mathrm{2}}\xspace}

\newcommand{\Dzero}        {\ensuremath{\mathrm{D^0}}\xspace}
\newcommand{\Dplus}        {\ensuremath{\mathrm{D^+}}\xspace}

\newcommand{\Dstar}        {\ensuremath{\mathrm{D^{*+}}}\xspace}

\newcommand{\Ds}           {\ensuremath{\mathrm{D_s^+}}\xspace}
 \newcommand{\Bs}           {\ensuremath{\mathrm{B_s^0}}\xspace}
 \newcommand{\Bplus}           {\ensuremath{\mathrm{B^+}}\xspace}
  \newcommand{\Bzero}           {\ensuremath{\mathrm{B^0}}\xspace}

\newcommand{\Lc}           {\ensuremath{\Lambda_\mathrm{c}^+}\xspace}
\newcommand{\Lcgeneric}           {\ensuremath{\Lambda_\mathrm{c}}\xspace}

\newcommand{\SigmacZeroPlus}           {\ensuremath{\Sigma_\mathrm{c}(2455)^{0,++}}\xspace}

\newcommand{\Sigmac}           {\ensuremath{\Sigma_\mathrm{c}^{0,+,++}}\xspace}
\newcommand{\Sigmacgeneric}           {\ensuremath{\Sigma_\mathrm{c}}\xspace}

\newcommand{\XicZero}      {\ensuremath{\Xi_\mathrm{c}^0}\xspace}
\newcommand{\XicPlus}      {\ensuremath{\Xi_\mathrm{c}^+}\xspace}
\newcommand{\XicPlusZero}  {\ensuremath{\Xi_\mathrm{c}^{0,+}}\xspace}
\newcommand{\Omegac}       {\ensuremath{\Omega_\mathrm{c}^0}\xspace}
\newcommand{\Jpsi}         {\ensuremath{\mathrm{J}/\psi}\xspace}
\newcommand{\UpsilonOneS}         {\ensuremath{\mathrm{\Upsilon (1S)}}\xspace}

\newcommand{\Lb}     {\ensuremath{\Lambda_\mathrm{b}^0}\xspace}

\newcommand{\herwig}    {\textsc{herwig}\xspace}
\newcommand{\pythiaeight}  {\textsc{pythia8}\xspace}

\newcommand{\pythia}       {\textsc{pythia}\xspace}

\title{Hadronization mechanisms (via heavy-flavour hadrons): Experiment}

\author*[a]{A. Rossi}

\affiliation[a]{INFN, sezione di Padova, \\
  via Marzolo 8, Padova, Italy}

\emailAdd{andrea.rossi@pd.infn.it}

\abstract{The formation of hadrons is a fundamental process in nature that can be investigated at particle colliders. Given their large mass, heavy quarks (charm and beauty) are produced only in initial hard-scatterings, prior to hadronisation, which determines instead the relative abundances and the kinematics of the various heavy-flavour hadron species. 
As several recent findings demonstrate, with \ee collisions as a "vacuum-like" reference at one extreme, and central AA as a dense, extended-size system characterised by flow and local equilibrium at the opposite extreme, different collision systems offer a lever arm that can be exploited to probe with a range of heavy-flavour hadron species the onset of various hadronisation processes. 
In these proceedings, a selection of the experimental results related to heavy-flavour hadronisation shown
for the first time at the Hard Probes 2023 conference is presented together with some of the most important ones of the last years. The focus is on open-heavy flavour measurements. The comparison with model predictions and connections among the results in \ee, proton--proton, proton--nucleus, nucleus--nucleus collisions are discussed. 
}
\FullConference{HardProbes2023\\
 26-31 March 2023\\
 Aschaffenburg, Germany\\}

\begin{document}
\maketitle

\section{Introduction}
The formation of hadrons is a fundamental process in nature that can be investigated at particle colliders. Given their large mass, heavy quarks (charm and beauty)
are produced mainly in high-$Q^2$ hard scatterings occurring between partons of the colliding nuclei, and, contrary to light quarks, their production in the soft processes
occurring during the hadronisation phase is suppressed. Therefore, heavy-quarks offer a specific perspective for investigating hadronisation and its modification
across collision systems, from \ee to proton--proton (pp) and nucleus--nucleus (AA) collisions, which differ in terms of ``size'', number of initial partonic interactions, density, kinematics, and relevant colour topologies connecting the produced quarks and gluons.

In recent years, the paradigm that heavy-quark hadronisation should proceed similarly in $\mathrm{e^{+}e^{-}}$ and pp collisions, which
motivated the usage of a factorisation approach for calculating charm- and beauty-hadron production
cross sections, has been severely questioned by the observation that charm and beauty baryon production relative to that of mesons
is larger in pp than in $\mathrm{e^{+}e^{-}}$ collisions. Tracing the modification of the hadronisation process
in different hadronic environments, from pp to AA collisions, and the possible emergence of quark coalescence as a
relevant hadronisation process already in small collision systems, is a major goal of current and future experiments. In these proceedings, some of the results shown
for the first time at the Hard Probes 2023 conference are presented together with some of the most important ones of the last years. The focus is on open heavy-flavour baryons and mesons: quarkonia and exotica are covered in~\cite{SmithHPProceedings} and~\cite{EscobedoHPProceedings}.

\begin{figure}[htb]
  \centering
  \includegraphics[width=0.52\textwidth]{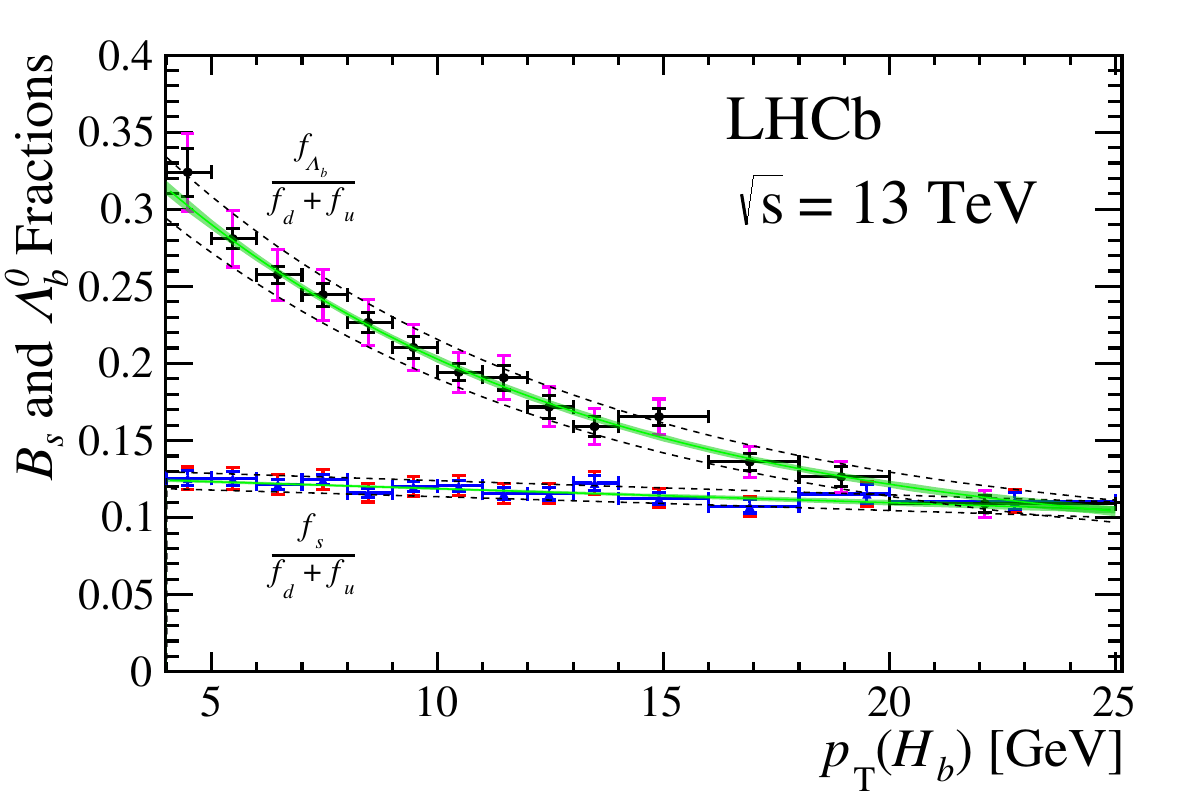}
\includegraphics[width=0.25\textheight]{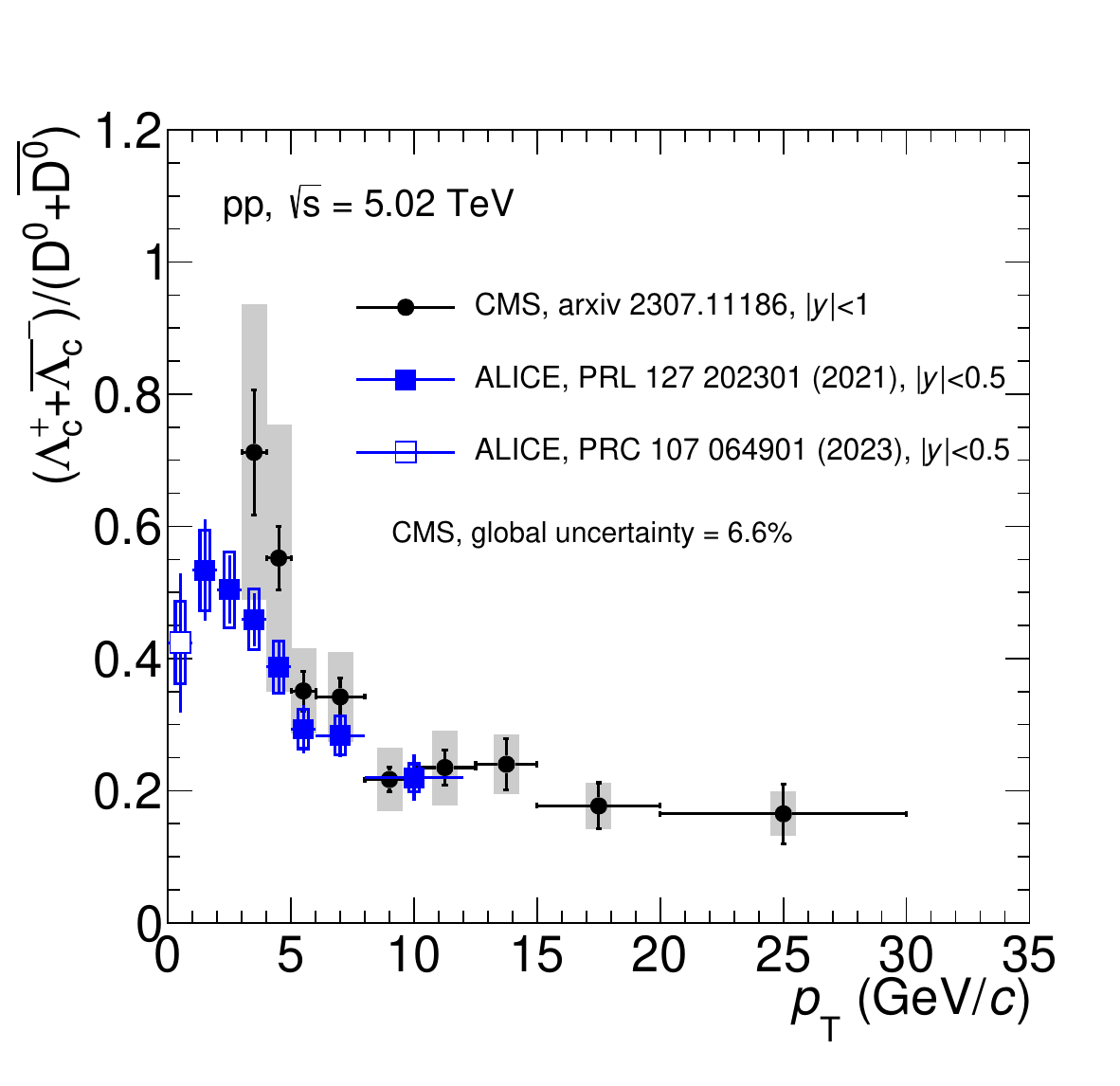}
\caption{Left: $\Lb/(\Bzero+\Bplus)$ and $\Bs/(\Bzero+\Bplus)$ ratio as a function of \pt measured by LHCb in $2<y<4.5$~\cite{LHCb:2019fns}. Right: $\Lc/\Dzero$ ratio as a function of \pt measured by ALICE~\cite{ALICE:2020wfu,ALICE:2022exq} and CMS at $\sqrt{s}=5.02$~\TeV~\cite{CMS:2023frs}.} 
\label{fig:LbLcPPaliceCMS}
\end{figure}
\section{Heavy-flavour baryon production in proton--proton collisions}
Already from the comparison of the baryon-to-meson \Lb/non-strange B cross-section ratio measured in \ppbar and \ee collisions at Tevatron and LEP, respectively, it emerged that the production of heavy-flavour baryon states could differ in hadronic with respect to \ee collisions~\cite{HFLAV:2016hnz}.  
At the LHC this ratio was measured with high precision by the LHCb Collaboration in the rapidity
interval $2<y<4.5$~\cite{LHCb:2019fns}. In Fig.~\ref{fig:LbLcPPaliceCMS}, the
ratio measured in pp collisions at $\sqrt{s}=13$~\TeV is reported as a function of \pt. In the same figure, the $\Bs/(\mathrm{B^{0}+B^{+}})$ ratio
is shown. A strong dependence on \pt, with an increase towards small \pt values, is visible for the $\Lb/(\mathrm{B^{0}+B^{+}})$ ratio. At LEP similar values, around 0.11, were measured for the two ratios~\cite{HFLAV:2016hnz}. The comparison indicates a
significant modification of the beauty-baryon production yield relative to B mesons from \ee to pp collisions.
A similar trend is observed in the charm sector: in the right panel of Fig.~\ref{fig:LbLcPPaliceCMS} the $\Lc/\Dzero$ ratios measured at midrapidity by ALICE~\cite{ALICE:2020wfu,ALICE:2022exq} and CMS~\cite{CMS:2023frs} in pp collisions at $\sqrt{s}=5.02$~\TeV are shown. 
The CMS and ALICE data are compatible within uncertainties. At high $\pt$ the ratio approaches the values measured in \ee collisions at LEP and B-factory energies as well as those measured in $\mathrm{e^\pm}p$ collisions at HERA~\cite{Lisovyi:2015uqa} (see Fig.~\ref{fig:FFpPbLcPtIntVsSyst}). The fact that in \ee collisions the same value is found at very different collision energies indicates
that the difference observed in pp collisions cannot be ascribed solely to the hard-scattering $Q^{2}$ and to the charm-quark energy. Moreover, the fact that
a similar difference is observed in the charm and beauty sector, and that the $\Bs/(\mathrm{B^{0}+B^{+}})$ ratio is very similar in pp and \ee collisions, suggests
that the effect is specific to baryons and cannot depend only on a kinematic scale defined by the quark or hadron masses. The similarity of the values measured at LEP and HERA
may suggest that the heavy-quark production process involved in the hard-scattering and the colour-topology of the hard-scattering products do not significantly
alter the hadrochemistry, though the argument cannot be pushed too much without a detailed analysis of the kinematic ranges involved.
This similarity was used to support the hypothesis of universality of fragmentation functions (FF)~\cite{ZEUS:2013fws}. Though not guaranteed
by QCD, especially at low \pt, it was often assumed as a reasonable one the hypothesis that the universality should extend to pp collisions~\cite{Cacciari:2001td,Cacciari:2005uk,Kniehl:2004fy,Kniehl:2012ti}. This assumption was corroborated by D-meson and B-meson data~\cite{Cacciari:2012ny} and by predictions of event generators like \pythia~\cite{Skands:2014pea} and \herwig~\cite{Bahr:2008pv}, in which hadronisation is implemented via string breaking and cluster formation, respectively, that did not indicate a significant modification of heavy-flavour particle ratios. 
\begin{figure}[htb]
  \centering
  \includegraphics[width=0.42\textwidth]{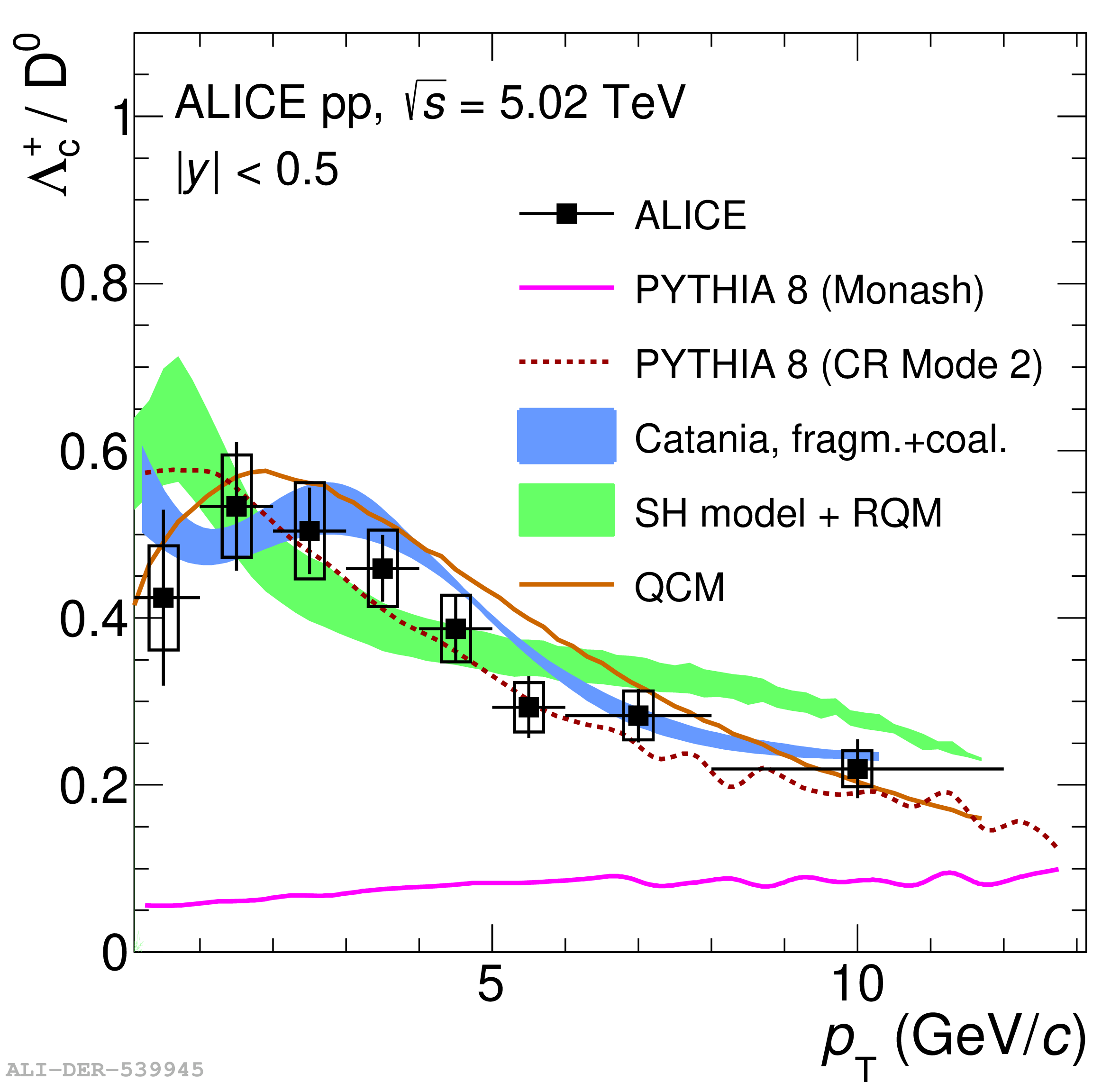}
\includegraphics[width=0.42\textwidth]{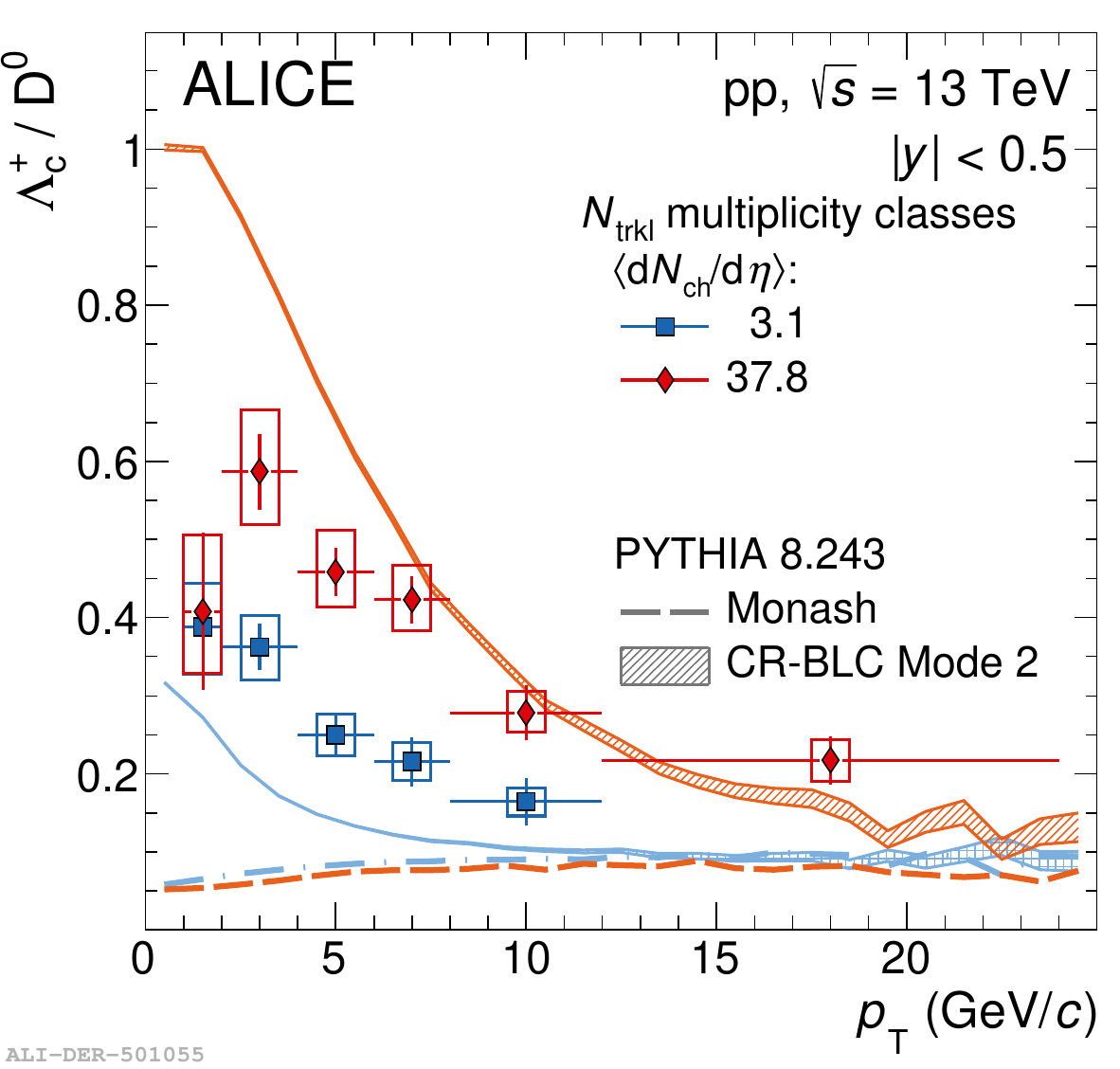}
\caption{Comparison of the \pt-differential $\Lc/\Dzero$ ratio measured by ALICE in pp collisions multiplicity-integrated (left)~\cite{ALICE:2020wfu,ALICE:2022exq} and at low and high multiplicity (right)~\cite{ALICE:2021npz} with model expectations (see text).} 
\label{fig:lcToDzeromodels}
\end{figure}
On the contrary, LHC data on heavy-flavour baryons largely violate this assumption at low \pt:
the observed \pt trend of the baryon-to-meson ratios clashes with that expected using \ee-based FF, and, as the detailed analysis carried out by ALICE by measuring the production of most ground-state charm baryons~\cite{ALICE:2021dhb} demonstrates, the fragmentation fraction of a heavy-quark to a given hadron species depend on the collision system. Therefore, universality does not extend to pp collisions. 
The enhancement of the baryon-to-meson ratios in pp collisions with respect to \ee collisions implies a modification of the hadronisation process. The above
considerations suggest that its cause must be searched ``outside the hard-scattering'' and ``standard'' jet fragmentation, possibly relating it to
differences in the collision-system properties that either modify the fragmentation process or favour different hadronisation mechanisms. The data also hint that at high \pt the ratios converge to the same values, recovering the validity of "in-vacuum" fragmentation and universality. If a kinematic scale had to be set to limit the range within which factorisation holds, it should be related to some system properties. Though a complete understanding is far from being reached, insight into modification of the hadronisation process can be obtained by comparing the measured ratios with theoretical models and by measuring their evolution with particle multiplicity.

Figure~\ref{fig:lcToDzeromodels} shows on the left the comparison of the $\Lc/\Dzero$ ratio with models as a function of \pt. The default version of \pythiaeight (Monash tune~\cite{Skands:2014pea}), in which string formation is implemented in the leading-colour approximation strongly underestimates the data. Within such approximation the beam-energy-fraction differential cross section of several charm baryons is reproduced in \ee collisions, as shown by Belle~\cite{Niiyama:2017wpp}. A tune of \pythiaeight (CR Mode 2~\cite{Christiansen:2015yqa}) with colour-reconnection (CR) implemented beyond leading colour (BLC) reproduces the pp data within uncertainties. In this model, when multiple parton interactions (MPI) are present, baryon production is enhanced thanks to the possibility of forming strings by colour-connecting in junction topologies quarks and gluons that are originated in different initial scatterings. 
The measured ratio is described within uncertainties also by the model of Ref.~\cite{He:2019tik} (labelled SH model+RQM in the figure), which can reproduce also the $\Lb/\mathrm{B^{-}}$ ratio measured by LHCb~\cite{He:2022tod}. In the model, the branching fractions of charm quarks to the various hadron species are determined by the thermal densities calculated with the Statistical Hadronisation Model (SHM~\cite{Andronic:2003zv}) and therefore depend only on the hadron mass and spin-degeneracy factor. The existence of a large set of excited charm-hadron states, richer of baryons than mesons, is assumed on the basis of the Relativistic Quark Model (RQM~\cite{Ebert:2011kk}) expectation. The large feed-down contribution from these excited states, most of which unobserved, is fundamental to reproduce the data. It was verified by ALICE that the feed-down from $\SigmacZeroPlus$ ground state is actually larger by a factor of about two with respect to \ee collisions and compatible with the model prediction~\cite{ALICE:2021rzj}. However, whether the unobserved excited states exist and are produced with thermal yields remain an open question. At the LHC, LHCb, ATLAS, and CMS observed a conspicuous number of charm and beauty states, many of which were not previously observed in \ee collisions either because energetically out of reach (as beauty baryons at B-factory energies), or, probably, because of a lack of statistics, which might even be related to lower production rates in \ee than hadronic collision systems. Nevertheless, the number of observed states (28 new baryons at the time of the conference~\cite{LHCb-FIGURE-2021-001-report}) is still far from the RQM expectations and claims for further experimental search and for an effort towards new spectroscopy and production measurements. A step in this direction was done by ALICE in the meson sector with preliminary measurements of the ratios $\mathrm{D_{s1}^{+}\cdot BR(D_{s1}^{+}\rightarrow \Dstar K^{0}_{s})/\Ds}$ and $\mathrm{D_{s2}^{*+}\cdot BR(D_{s2}^{*+}\rightarrow \Dplus K^{0}_{s})/\Ds}$ in $2<\pt<24$~\GeVc~\cite{SPolitanoHPProceedings}, which result independent on multiplicity and compatible with SHM expectations both with and without considering unobserved excited states, a conclusion to be revised after the expected reduction of the experimental uncertainties with the analysis of LHC Run-3 data.
In the SHM approach, differently from the \pythiaeight case, there is not any modelling of the hadronisation process. In the Catania model~\cite{Minissale:2020bif} charm quarks can hadronise via \enquote{vacuum}-like fragmentation as well as recombine (coalesce) with surrounding light quarks from the underlying event, which is described as a quark-gluon plasma also in pp collisions. The probabilities that three (two) quarks form a given baryon (meson) are calculated by means of the Wigner formalism and depend on the hadron wave function and the phase-space distributions of the heavy and light quarks at a given hadronisation temperature at which coalescence occurs "suddenly"~\cite{Fries:2008hs,Plumari:2017ntm}. The normalisation is such that at $\pt=0$ all hadrons come from coalescence, while fragmentation takes over with increasing \pt. The Catania model well reproduces the $\Lc/\Dzero$ ratio. 
In the quark-(re)combination mechanism (QCM~\cite{Song:2018tpv}) charm quarks form hadrons by combining with equal-velocity light quarks. In this model, the relative abundances of the different baryon species are fixed by thermal weights, though the total charm baryon-to-meson ratio is constrained to the $\Lc/\Dzero$ ratio measured by ALICE in pp collisions at $\sqrt{s}=7$~\TeV. The \pt trend of the ratio is well described by the model.  
\begin{figure}[htb]
  \centering
  \includegraphics[width=0.32\textwidth]{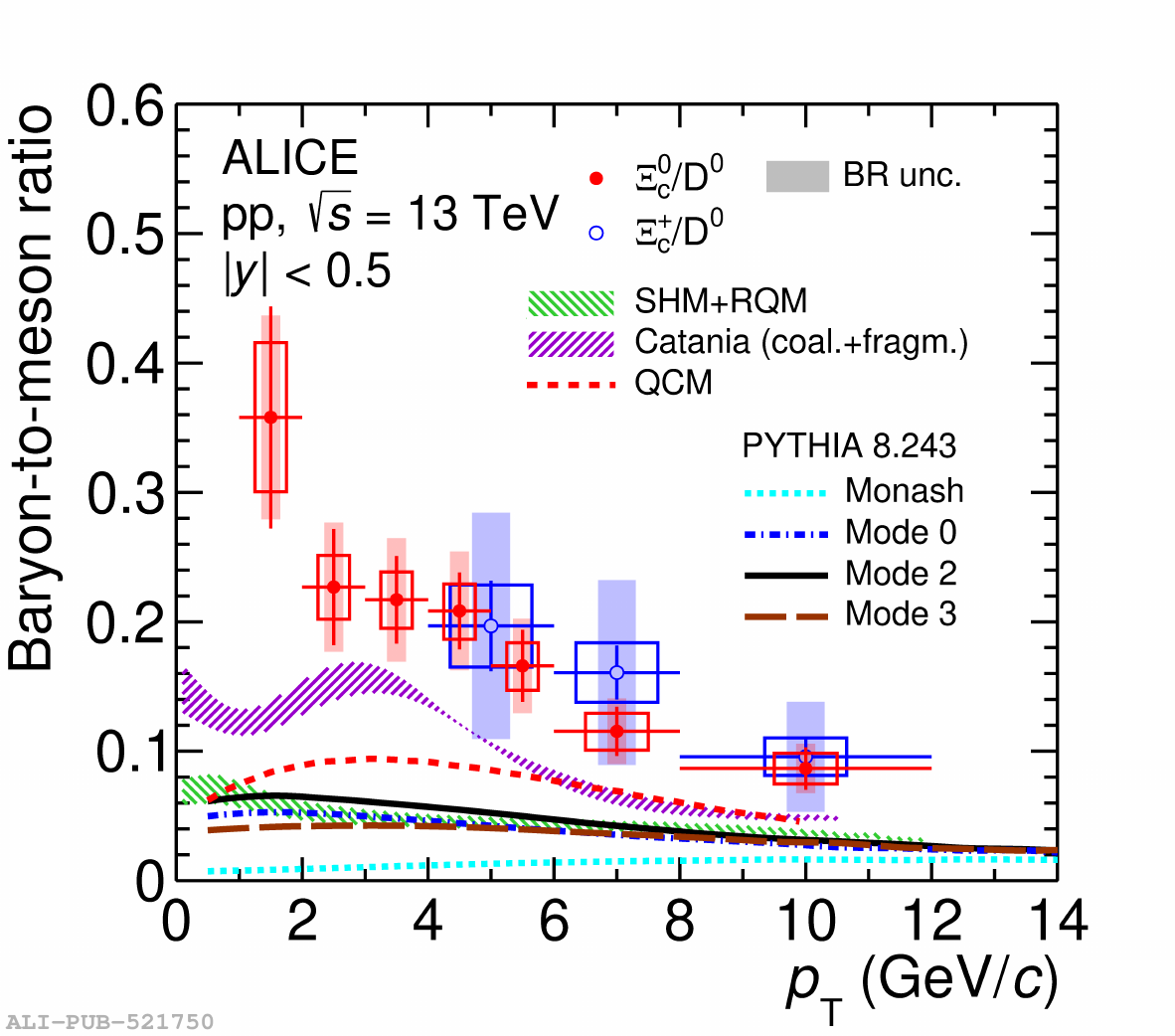}
\includegraphics[width=0.32\textwidth]{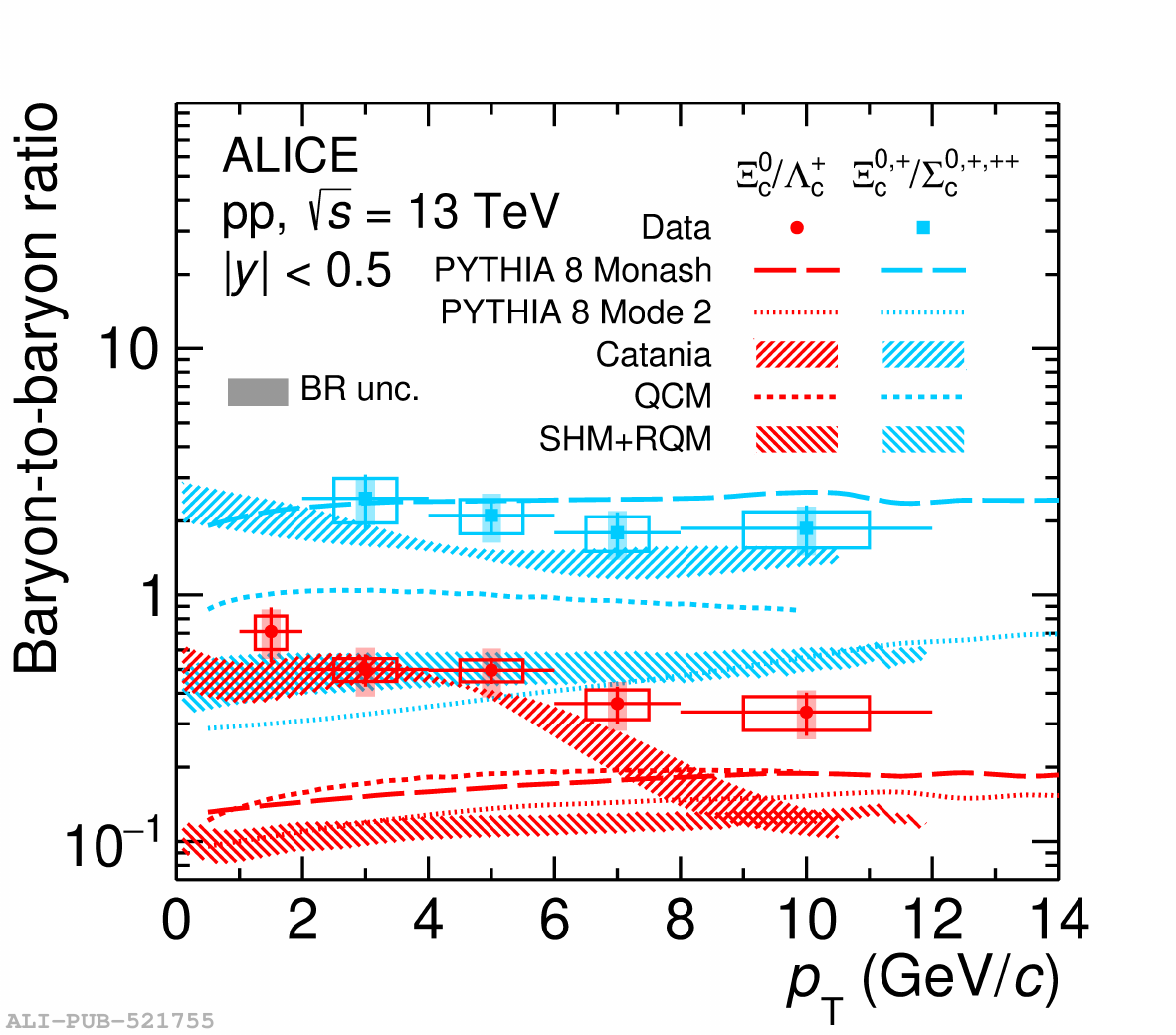}
\includegraphics[width=0.26\textwidth]{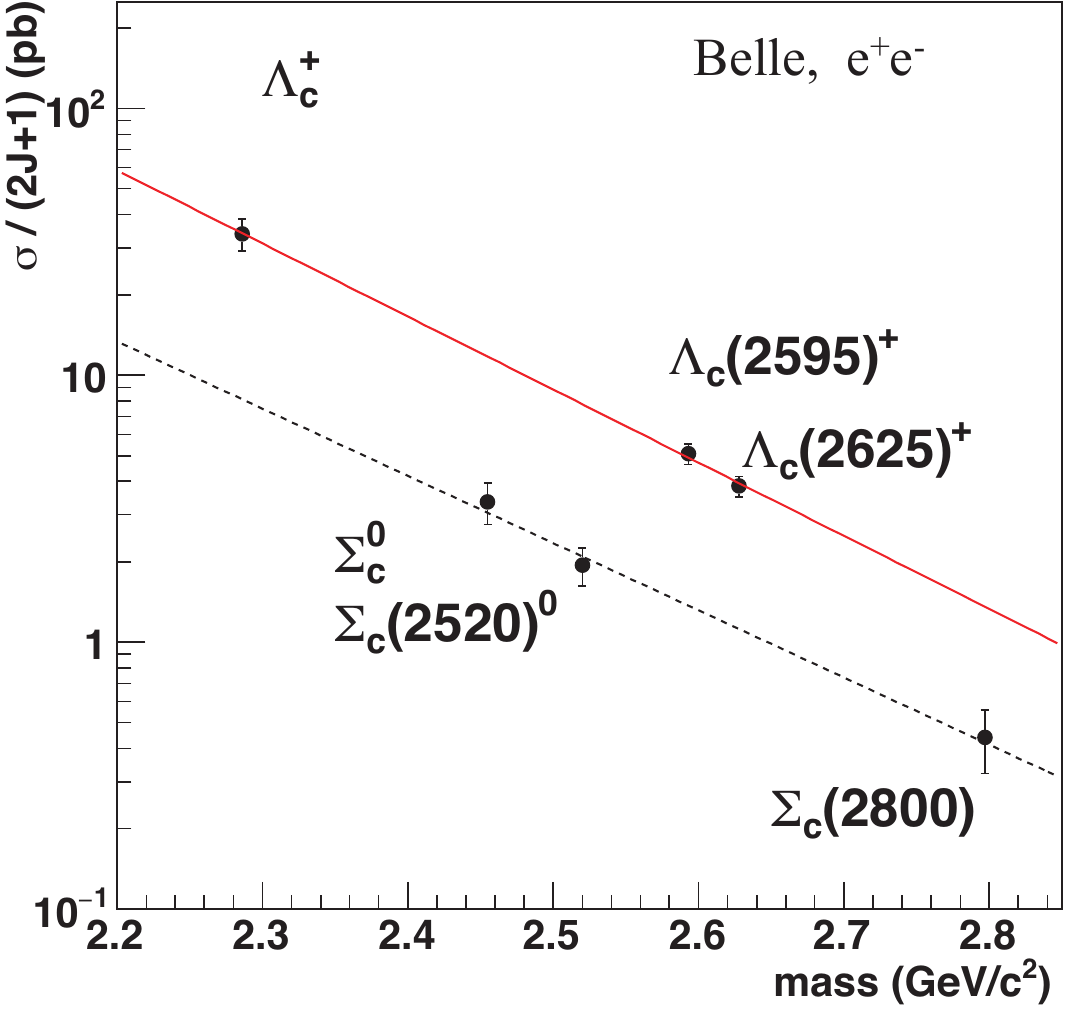}
\caption{Comparison of \pt-differential $\XicPlusZero/\Dzero$ (left), $\XicPlusZero/\Sigmac$ and $\XicZero/\Lc$ (middle) ratios with model expectations in pp collisions~\cite{ALICE:2021bli,ALICE:2021rzj}. Right: cross sections of direct production of \Lcgeneric and \Sigmacgeneric states scaled by the spin-degeneracy factor measured by Belle in \ee collisions~\cite{Niiyama:2017wpp},  as a function of particle mass. }
\label{fig:XicScpp}
\end{figure}
\subsection{Strangeness and diquarks: \XicPlusZero, \Omegac, and \SigmacZeroPlus}
The ALICE experiment measured at midrapidity the \pt-differential production cross section of $\XicPlusZero$ baryons~\cite{Acharya:2021dsq,ALICE:2021bli}, $\SigmacZeroPlus$~\cite{ALICE:2021rzj}, and the cross section times branching ratio (BR) of the \Omegac baryon~\cite{ALICE:2022cop}. As shown in Fig.~\ref{fig:XicScpp}, the $\XicZero/\Dzero$ ratio shows a similar tendency than the $\Lc/\Dzero$ ratio to decrease with \pt and, at low \pt, is higher than the \ee value of about 0.02, which can be estimated from $\XicPlus/\Lc$ and $\Lc/\Dzero$ data~\cite{Niiyama:2017wpp,Lisovyi:2015uqa,ParticleDataGroup:2022pth}. 
All models tend to underestimate the data, with the Catania one predicting the values closer to the measured ratio. In the \Omegac case~\cite{ALICE:2022cop}, the limited knowledge of the branching ratio of the $\Omegac\rightarrow\Omega^{-}\pi^{+}$ decay channel prevent firm conclusions. However, even more than for the \XicPlusZero case, the model expectations differ significantly, by orders of magnitude, highlighting the sensitivity of this observable to the details of the hadronisation mechanism. 
The failure of most models in reproducing the $\XicPlusZero$ and $\Omegac$ results may induce to focus on strangeness, which can reasonably demand specific treatment considering the large strange-quark mass and strangeness canonical suppression~\cite{Tounsi:2001ck}. However, the $\Bs/(\Bzero+\Bplus)$ and $\Ds/\Dzero$~\cite{ALICE:2021mgk} ratios do not show strong modifications (if any) with respect to \ee collisions. Looking at the middle panel of Fig.~\ref{fig:XicScpp}, it is remarkable to note that the Monash tune of \pythiaeight can reproduce within uncertainties the $\XicZero/\Sigmac$ ratio, despite the fact that it severely underpredicts the $\XicZero/\Dzero$ and $\SigmacZeroPlus/\Dzero$ ratios, which are both enhanced more than the $\Lc/\Dzero$ in pp with respect to \ee collisions. In \ee collisions the production of $\mathrm{\Sigma_{c}}$ states is suppressed with respect to that of $\mathrm{\Lambda_{c}}$ states~\cite{Niiyama:2017wpp}, as shown in the right panel of Fig.~\ref{fig:XicScpp}. In the Lund string model charm baryons are formed in \ee collisions by attaching a light-quark diquark to the charm quark, which is a string end point. Diquarks are produced in pair with anti-diquarks in string breaking via the Schwinger mechanism. For $\mathrm{\Sigma_{c}}$ states (ud,uu,dd) diquarks with spin (S) and isospin (I) (S=1,I=1) are needed, while (ud) diquarks with (S=0,I=0) are required for $\mathrm{\Lambda_{c}}$ states. Given that the former are heavier, their production via the Schwinger mechanism is suppressed. In most calculations~\cite{Barabanov:2020jvn}, a similar mass is obtained or assumed for S=1 (ud,uu,dd) diquarks and S=0 (ds) diquarks, contained in $\XicZero$. The fact that Monash tune can reproduce the $\XicZero/\Sigmac$ ratio in pp collisions could be a coincidence (the predicted value remains to be explained along with its dependence on the two particle masses, almost identical, and the assumed feed-down contributions) but it may also signal the removal of a similar suppression factor affecting both $\XicZero$ and $\SigmacZeroPlus$ production in \ee collisions. This could suggest that light diquarks are more easily produced in pp than \ee collisions via other processes than the Schwinger mechanism, or that charm-baryon formation does not necessitate their presence. In the \pythiaeight CR-BLC model, heavy-flavour baryon production is increased by the formation of heavy-light diquarks in junction CR topologies, while in coalescence models already existing quarks recombine. 
Clarifying the role of light and heavy diquarks will be important for understanding hadronisation in pp as well as in heavy-ion collisions, where diquarks could also be effective degrees of freedom of the system close to the phase transition~\cite{Beraudo:2023fpq}.
\begin{figure}[ht!]
    \begin{center}
    \includegraphics[width = 0.35\textwidth]{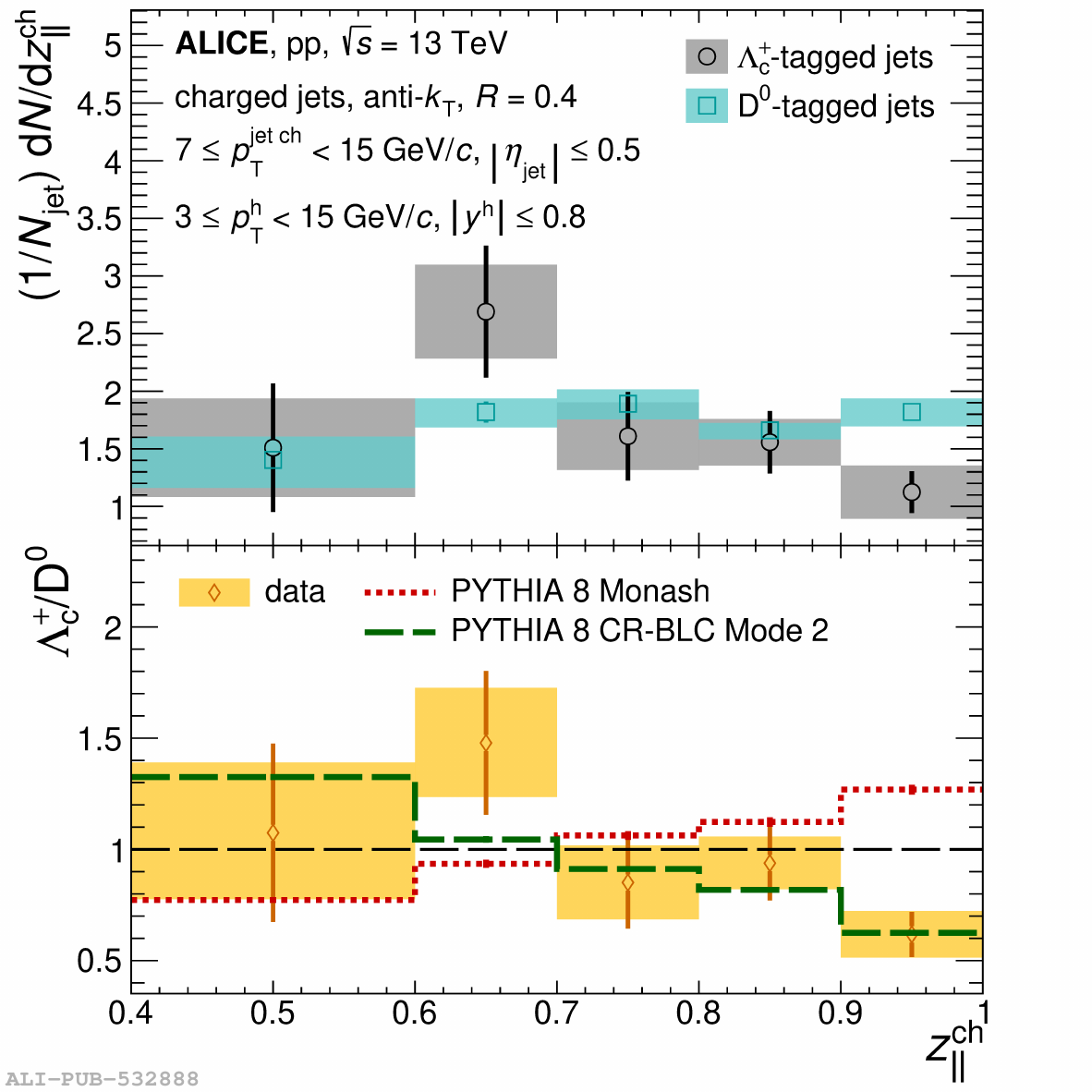}
   \includegraphics[width = 0.4\textwidth]{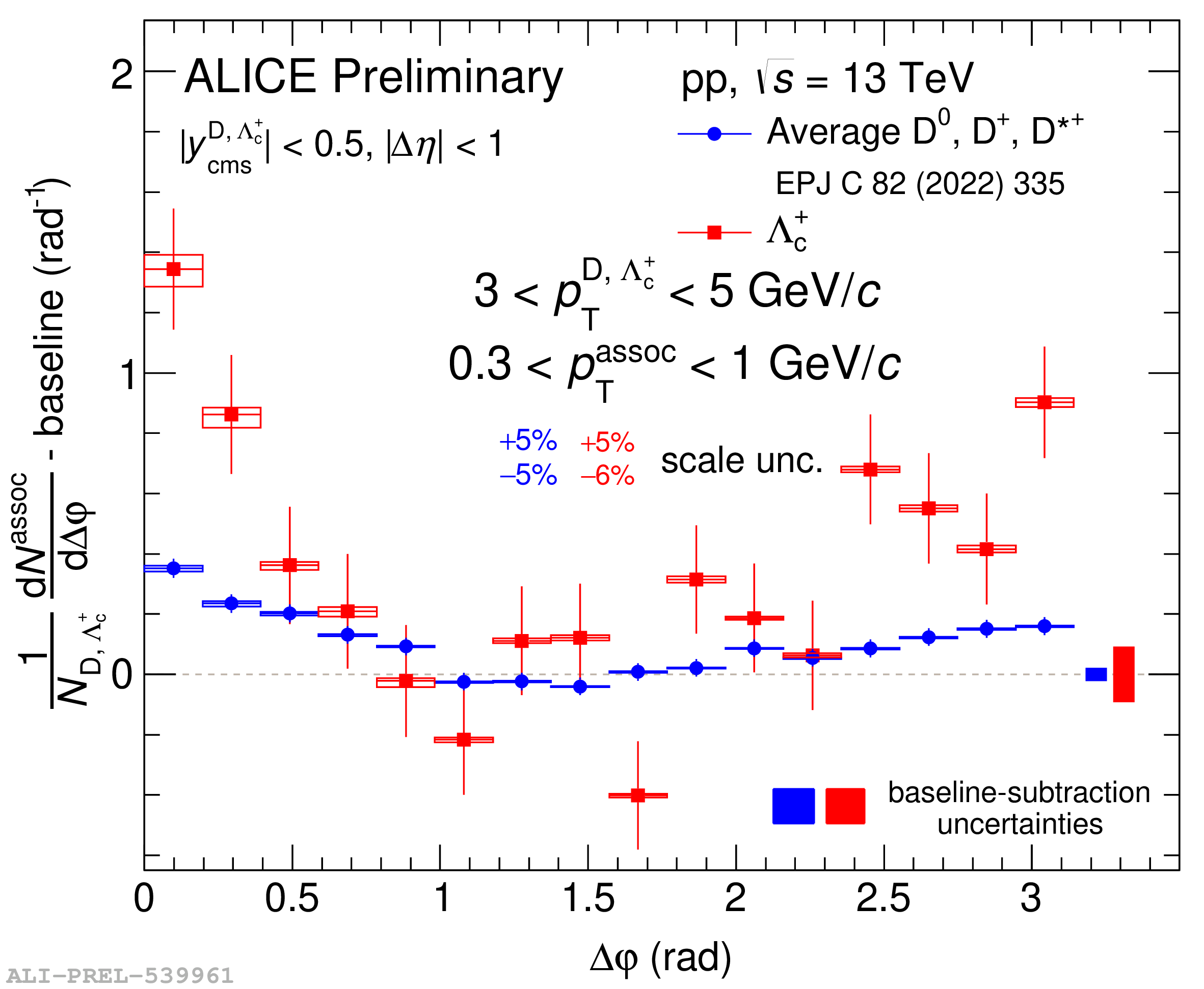} 
    \end{center}
    \caption{Left: comparison (top panel) and ratio (bottom panel) of the distributions of the \Lc and \Dzero longitudinal track-based jet-momentum fraction ($\zjet$)~\cite{ALICE:2023jgm}. 
    Right: comparison of the $\Lc$-charged hadrons and $\Dzero$-charged hadrons azimuthal correlation functions after baseline subtraction~\cite{APalascianoHPProceedings}.}
    \label{fig:LcAndDtaggedJets}
\end{figure}
\subsection{Jets and correlations}
Additional information for investigating the heavy-flavour hadronisation process is provided by beauty- and charm-tagged jets especially when jets are tagged by the presence of a fully reconstructed heavy-flavour hadron, which allows to compare the hadron and jet momentum vectors. As shown in Fig.~\ref{fig:LcAndDtaggedJets} (left panel), there is an indication that, at low jet \pt, the distribution of the fraction ($\zjet$) of jet momentum carried by the heavy-flavour hadron along the jet axis is different for $\Dzero$- and $\Lc$-tagged track-based jets~\cite{ALICE:2023jgm}. 
Though the uncertainties prevent firm conclusions, the data suggest that $\Lc$ baryons carry a smaller fraction of jet momentum along the jet axis with respect to D mesons. A softer $\zjet$ distribution can be related to a smaller hadron jet-energy fraction or a smaller alignment of the hadron and jet momentum vectors, both aspects sensitive to a possible interplay with the underlying event. The $\Lc$-jet measured trend contrasts with that expected by the default version of \pythiaeight (Monash tune), which reasonably describes D-meson tagged jets~\cite{ALICE:2022mur} and reflects the tendency in jet fragmentation for heavier particles to carry a larger fraction of jet momentum, while it agrees with that expected by the tune of \pythiaeight with CR BLC. 
\begin{figure}[!t]
\centering
\includegraphics[width=0.365\textwidth]{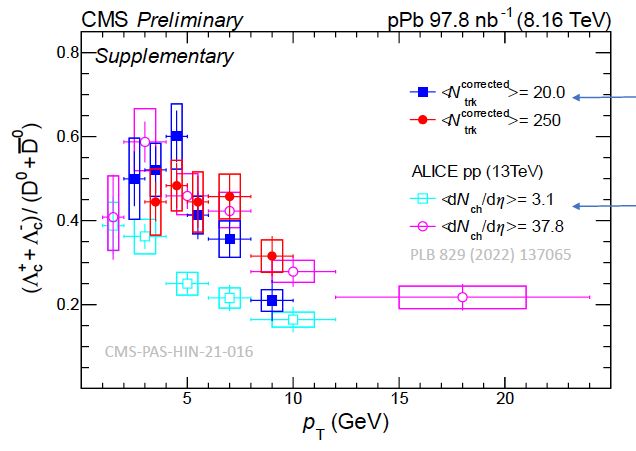}
\includegraphics[width=0.26\textwidth]{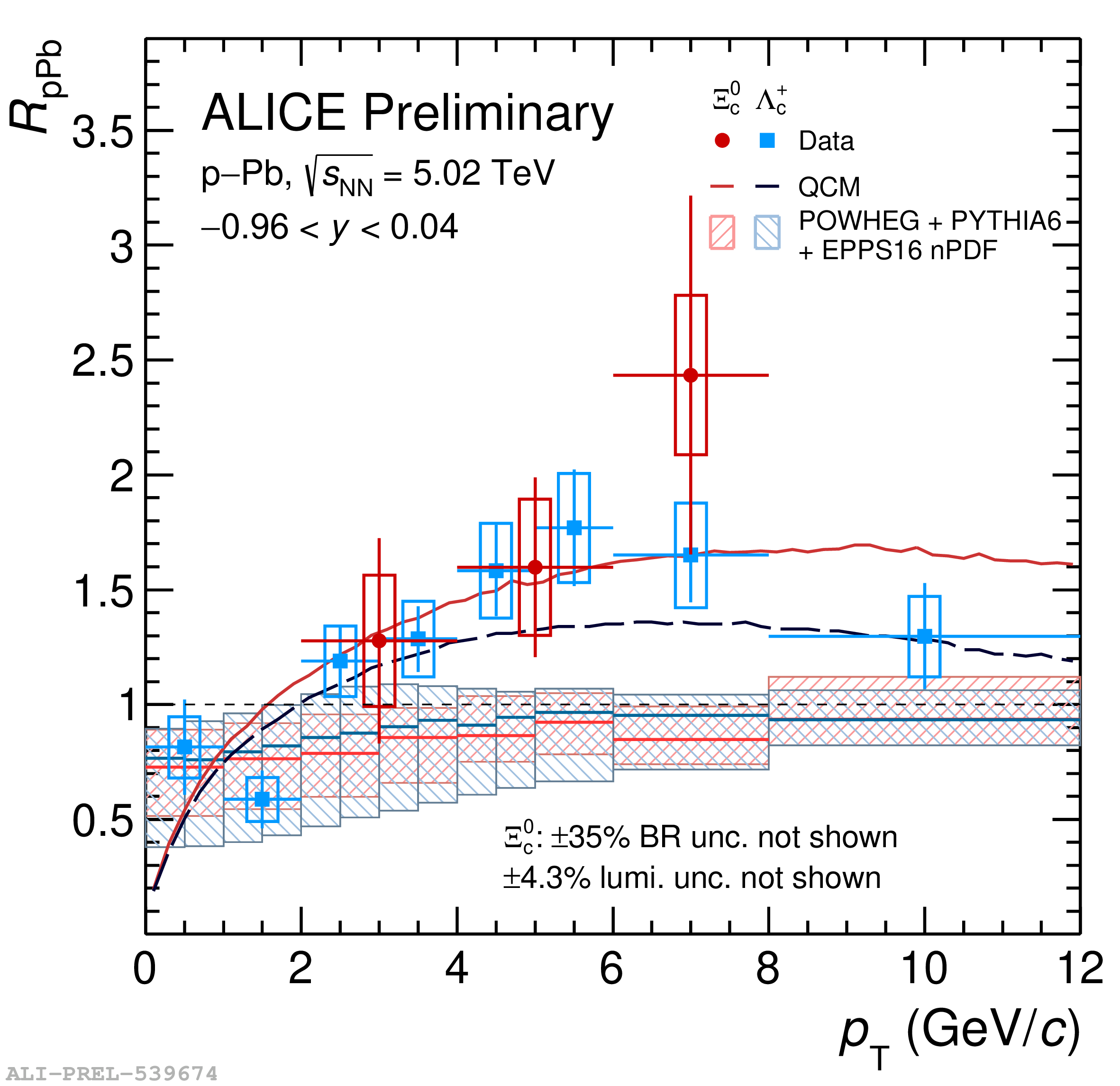}
\includegraphics[width=0.32\textwidth]{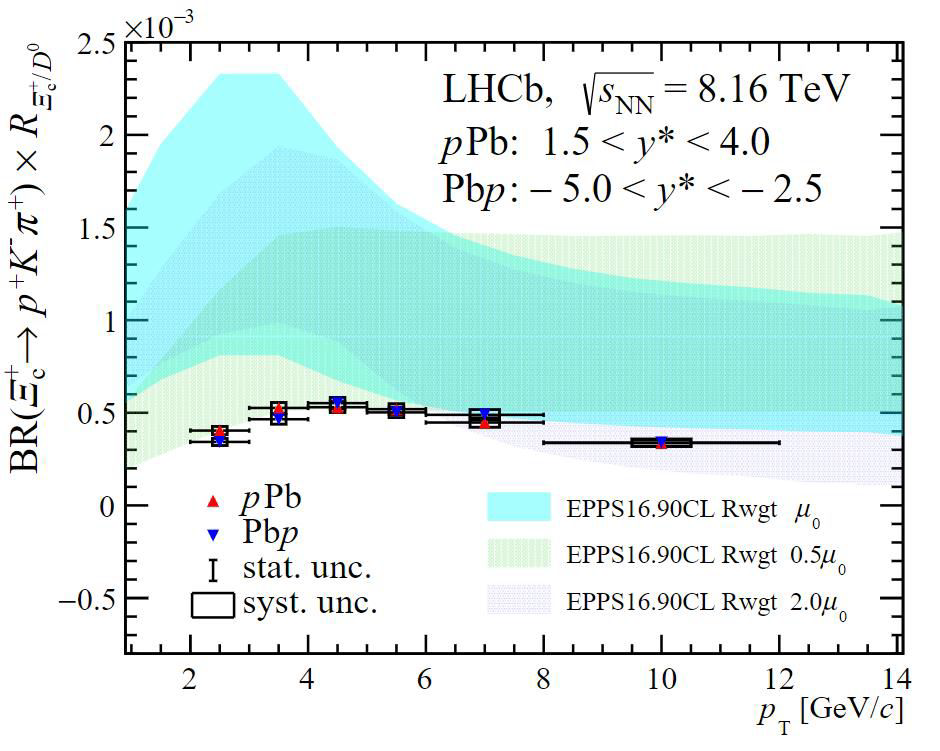}
\caption{Left: $\Lc/\Dzero$ ratio as a function of \pt measured in different multiplicity intervals by ALICE in pp collisions~\cite{ALICE:2021npz} and by CMS in p--Pb collisions~\cite{YZhangHPProceedings}. Middle: nuclear modification factor of $\Lc$ and $\XicZero$ production as a function of \pt in p--Pb collisions compared with model predictions ~\cite{AKalteyerHPProceedings}. Right: $\XicPlus/\Dzero$ ratio at backward and forward rapidity measured by LHCb in p--Pb collisions~\cite{RLitvinovHPProceedings}.} 
\label{fig:LcXicpPb}
\end{figure}
A related, study is that of the azimuthal correlation of \Lc baryons with charged particles produced in the event. As shown in the right panel of the Fig.~\ref{fig:LcAndDtaggedJets}, a new preliminary measurement, performed by ALICE~\cite{APalascianoHPProceedings}, indicates that the yields of low-\pt associated particles are large, both in the near-side and and in the direction opposite to the \Lc (away-side jet region), and resemble those observed only at much higher D-meson \pt ($\pt\gtrsim 8$~\GeVc) in correlations of D mesons with charged particles. The naïve expectation would be that, a part from the initial-parton energy correlation, the fragmentation (parton-shower) and hadronisation of the near- and away-side jets should be almost independent. It is thus surprising that the larger near-side yield, which is consistent with the softer jet fragmentation suggested by the $\zjet$ analysis, is accompanied by a difference in the away-side jet. Taken together, the $\zjet$ and azimuthal correlation data may suggest that, with respect to D mesons, low \pt $\Lc$ are produced on average in jets with higher-\pt and softer fragmentation, with a sharing of the jet energy among several soft particles. Whether this could be due to an interplay with the underlying event is not clear, though possible, considering also the multiplicity dependence of $\Lc/\Dzero$ ratio (see next section). More precise measurements and a detailed analysis of the jet radial profiles and substructure and of the correlation-function widths could in the future shed light on this intriguing observations. It would also be important to understand what hadron-formation via coalescence implies for these observables. 
\begin{figure}[htb]
\centering
\includegraphics[width=0.39\textwidth]{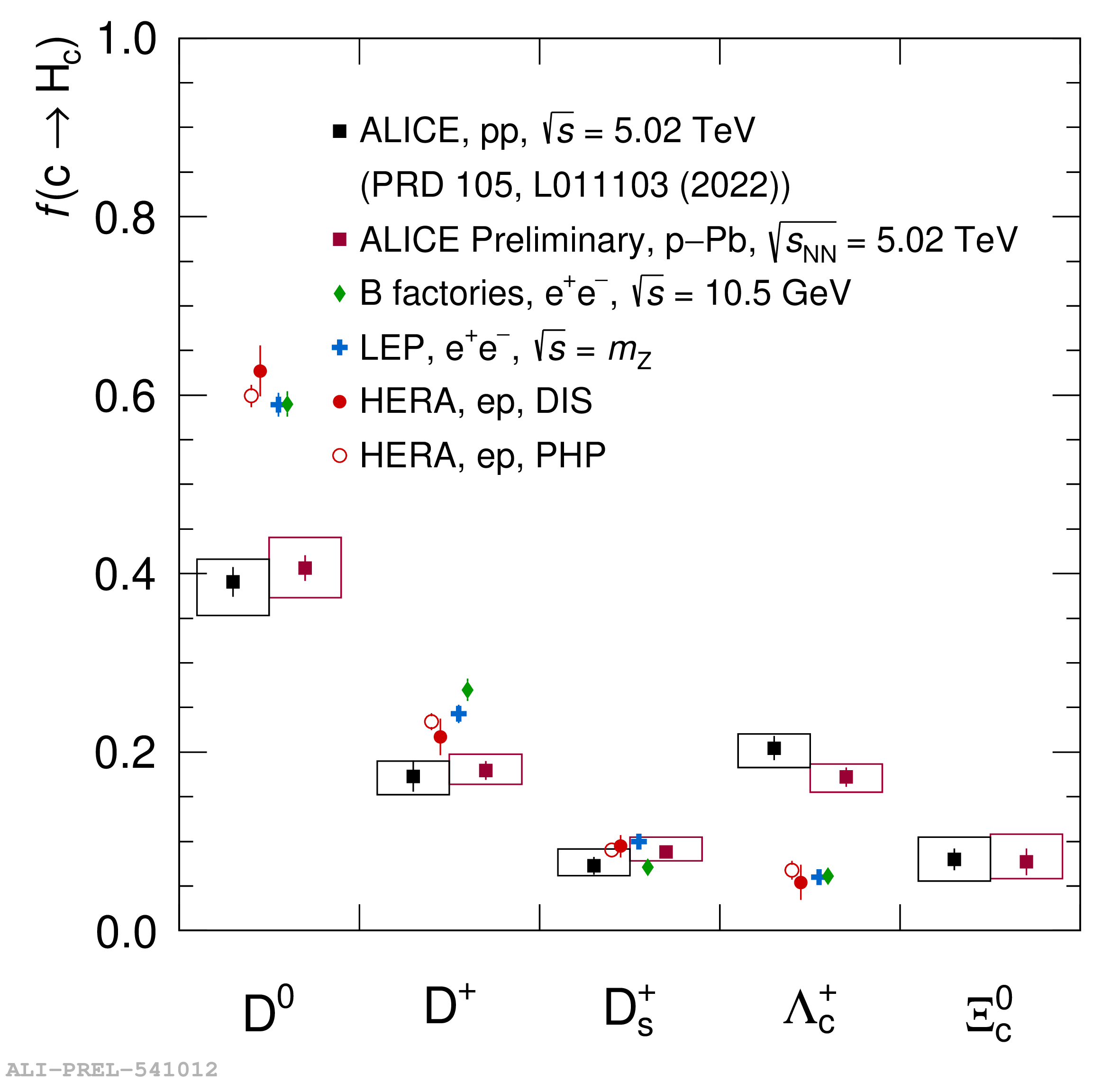}
\includegraphics[width=0.43\textwidth]{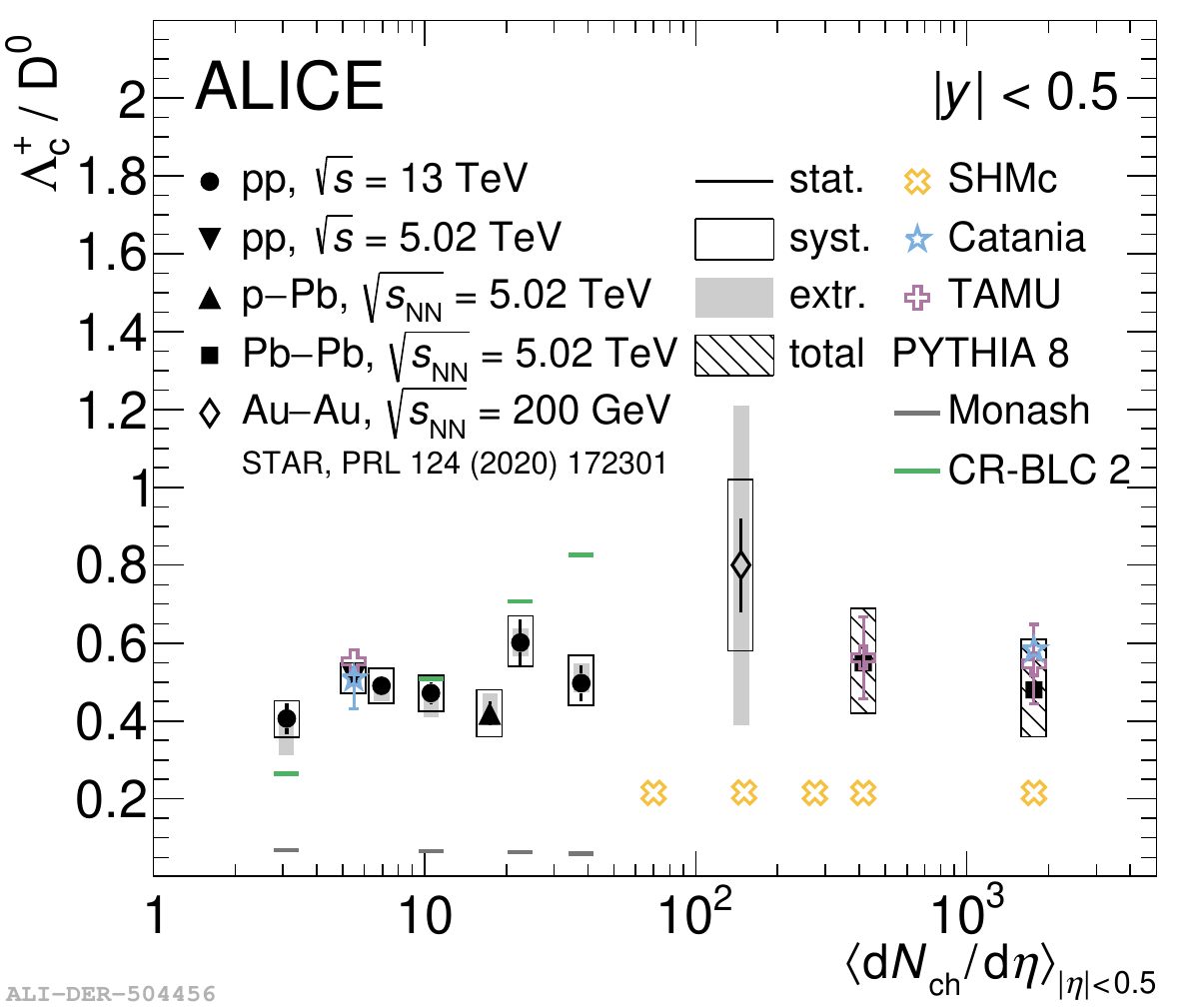}
\caption{Left: comparison of charm fragmentation fractions in pp, p--Pb, \ee, and ep collisions~\cite{AKalteyerHPProceedings}. Right: \pt-integrated $\Lc/\Dzero$ ratio at midrapidity as a function of charged-particle multiplicity~\cite{ALICE:2021npz,ALICE:2021bib}.} 
\label{fig:FFpPbLcPtIntVsSyst}
\end{figure}
\section{Dependence on event activity: pp and p-Pb collisions}
In the right panel of Fig.~\ref{fig:lcToDzeromodels} it is shown that for $\pt>2$~\GeVc the $\Lc/\Dzero$ ratio increases from low to high multiplicity in pp collisions, a trend qualitatively described by \pythiaeight with CR BLC. In this tune, the baryon-to-meson ratio increases naturally with the number of MPI, thus with increasing particle multiplicity. However, the observed difference is smaller than that expected by the model, which also predicts a continuous rise of the ratio down to \pt=0 which is not favoured by the high-multiplicity data. At forward rapidity, the LHCb Collaboration observed a dependence on multiplicity, estimated with VELO tracks, of the $\Bs/\Bzero$ ratio stronger than that expected by \pythiaeight both with and without colour reconnection~\cite{LHCb:2022syj,CGuHPProceedings}. This effect is not visible when the multiplicity is sampled far from the rapidity interval in which $\Bs$ and $\Bzero$ are reconstructed. 

Extending the analysis of baryon-to-meson ratios to p--Pb collisions, one can note in Fig.~\ref{fig:LcXicpPb} that a preliminary study by CMS shows that the \pt-differential $\Lc/\Dzero$ ratios in low- and high-multiplicity p--Pb collisions are similar to that measured in high-multiplicity pp collisions, "isolating" the one measured in low-multiplicity pp collisions. The modification from (multiplicity-integrated) pp to p--Pb collisions is reproduced by the QCM model~\cite{Song:2015ykw}. In p--Pb collisions, the absence of an evolution from low to high multiplicity breaks the similarity of $\Lc/\Dzero$ and $\Lambda/\mathrm{K^{0}_{s}}$ ratios observed by ALICE in pp collisions, also as a function of multiplicity~\cite{ALICE:2021npz}, and, within larger uncertainties, in multiplicity-integrated p--Pb collisions~\cite{ALICE:2020wfu}. For the first time, the \XicZero cross section and $\XicPlus$ cross section times BR($\XicPlus\rightarrow \mathrm{pK^{-}\pi^{+}}$) were measured in p--Pb collisions, by ALICE~\cite{AKalteyerHPProceedings} and LHCb~\cite{RLitvinovHPProceedings}, respectively. A hint that in the interval $2<\pt<8$~\GeVc the $\XicZero$ nuclear modification factor ($\RpPb$) is larger than unity and rises with \pt can be seen in the middle panel of Fig.~\ref{fig:LcXicpPb}. The comparison with $\Lc$ $\RpPb$ suggests a similar modification from pp to p--Pb of the \pt-differential cross sections of the two baryons, which is described within uncertainties by the QCM model, and is not present for D mesons. 
No significant difference is observed by LHCb between the $\XicZero/\Dzero$ ratio at forward and backward rapidity, as shown in the right panel of the same figure. 

As shown in Fig.~\ref{fig:FFpPbLcPtIntVsSyst} (left panel), the charm fragmentation fractions in pp and p--Pb collisions are compatible within uncertainties, both significantly differing from those in \ee and $\mathrm{e^{\pm}p}$ collisions. Contrary to the \pt-differential ratio, the ratio of the \pt-integrated $\Lc$ and $\Dzero$ spectra (right panel) remains the same, within uncertainties, from low-multiplicity pp up to the multiplicity of central Pb--Pb collisions, including STAR result in Au--Au collisions at $\snn=200$~\GeV~\cite{STAR:2019ank}. The Catania~\cite{Plumari:2017ntm} and TAMU~\cite{He:2019vgs} models can reproduce the data, the SHMc~\cite{Andronic:2021erx} without inclusion of extra excited states can reproduce only the trend, while \pythiaeight expects a rise of the ratio with increasing multiplicity in pp collisions which is disfavoured by the data. 
It must be noted that the data uncertainty in the lowest-\pt bins are typically large and more precise measurements down to $\pt=0$ are needed for a conclusive assessment of the \pt-integrated trends. It would also be important to test whether at lower multiplicities the $\Lc/\Dzero$ ratio tends to the \ee values. 
\subsection{Rapidity dependence: an open point}
While the yield ratio of non-prompt $\Lc$ to non-prompt $\Dzero$ measured at midapidity by ALICE~\cite{AKalteyerHPProceedings} in pp collisions is consistent with expectations based on LHCb $\Lb/(B^{0}+B^{+})$ at forward rapidity, the $\Lc/\Dzero$ ratios measured at midrapidity tends to be higher than those at forward rapidity in both pp and p--Pb collisions. Considering available measurements of the BR($\XicPlus\rightarrow \mathrm{pK^{-}\pi^{+}}$)~\cite{ParticleDataGroup:2022pth}, this trend is present also for $\XicPlusZero/\Dzero$ in p--Pb collisions. Hopefully, the rapidity dependence of the baryon-to-meson ratios will be clarified with new measurements in next years. 
\begin{figure}[t!]
\centering
\includegraphics[width=0.33\textwidth]{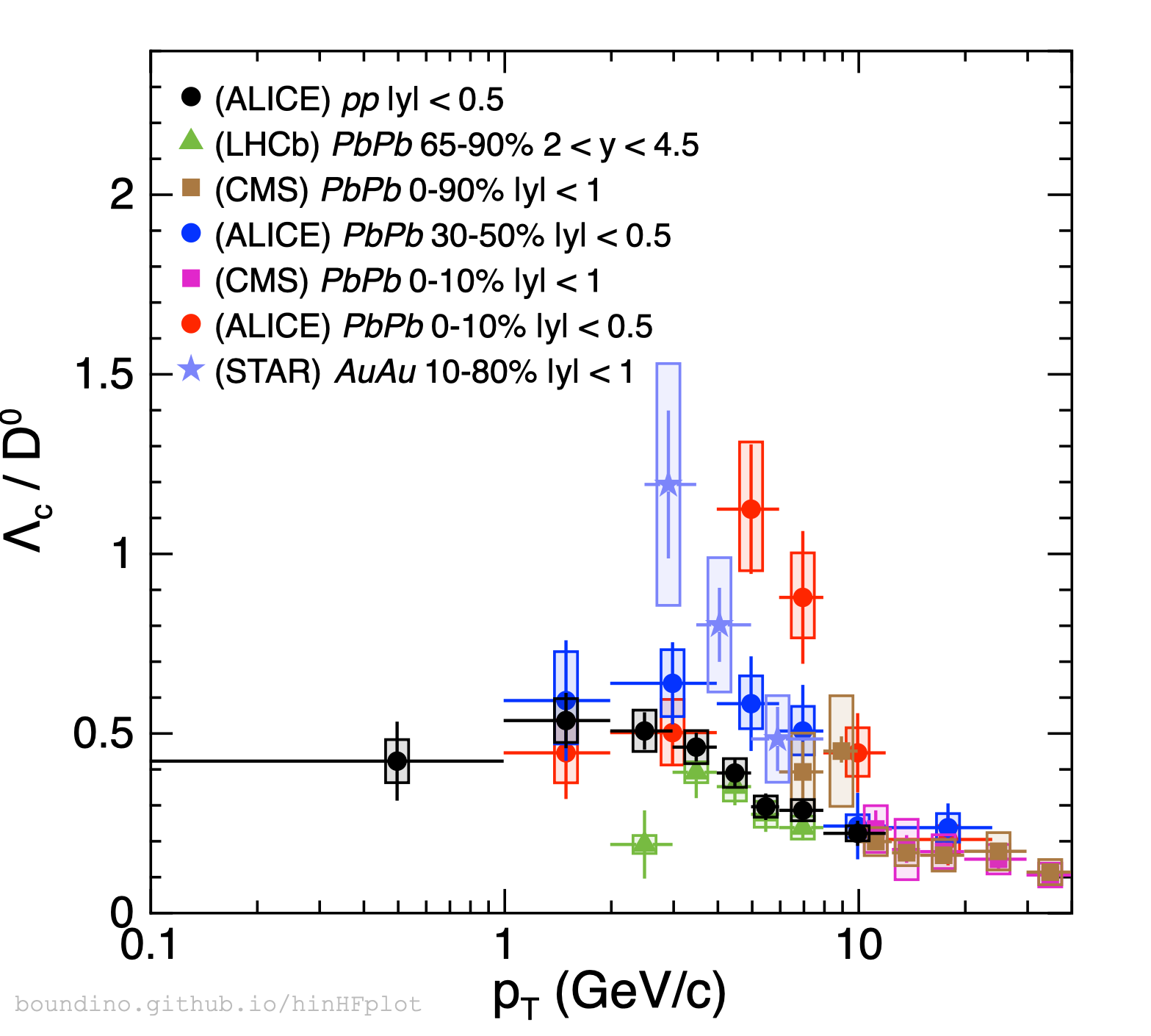}
\includegraphics[width=0.3\textwidth]{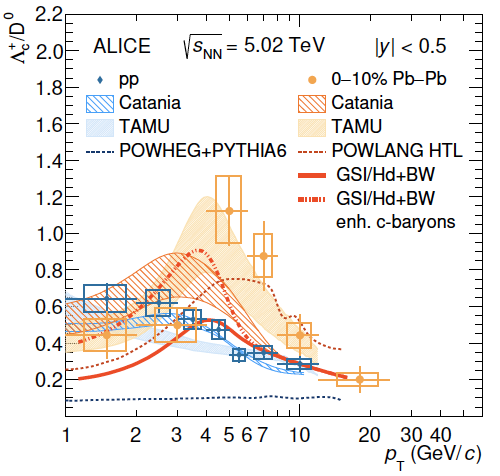}
\includegraphics[width=0.3\textwidth]{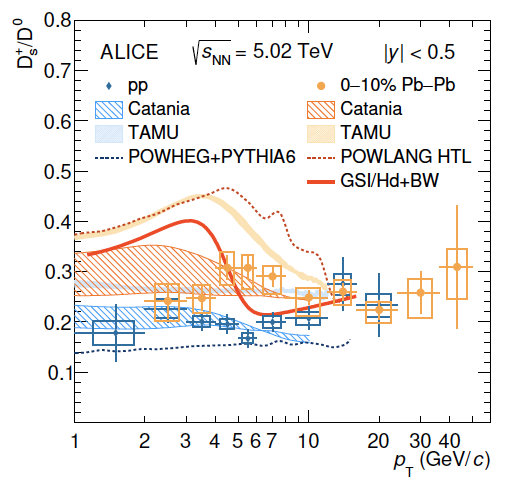}
\caption{Left: comparison of the $\Lc/\Dzero$ ratios measured as a function of \pt in pp~\cite{ALICE:2020wfu}, Pb--Pb~\cite{ALICE:2021bib,YZhangHPProceedings,CMS:2023frs,LHCb:2022ddg}, and Au--Au~\cite{STAR:2019ank} collisions. Middle, right ( from~\cite{ALICE:2022wpn}): comparison of $\Lc/\Dzero$ (middle) and $\Ds/\Dzero$ (right) ratios with model predictions~\cite{Minissale:2020bif,Plumari:2017ntm,Andronic:2021erx,He:2019tik,He:2019vgs,Beraudo:2022dpz} in pp and 0--10\% central Pb--Pb collisions.} 
\label{fig:LcPbPbAllexpModels}
\end{figure}
\section{Nucleus-nucleus collisions}
While, as discussed in the previous section, the \pt-integrated $\Lc/\Dzero$ ratio does not change significantly from pp to Pb--Pb collisions, it evolves substantially as a function of \pt with increasing centrality, as shown in Fig.~\ref{fig:LcPbPbAllexpModels} (left panel), where the Pb--Pb ALICE~\cite{ALICE:2021bib}, CMS~\cite{YZhangHPProceedings}, and LHCb~\cite{LHCb:2022ddg} results are reported and compared to ALICE pp data as well as to the measurement in Au--Au by STAR~\cite{STAR:2019ank}. At $\pt>20$~\GeVc the ratio is consistent with that in \ee collisions. At low-intermediate \pt a \enquote{radial-flow}-like peak emerges, possibly shifted at lower \pt at RHIC with respect to LHC energy. It most likely derives from hadronisation via coalescence, reinforced by space-momentum correlations (SMC), as the comparison with transport models (right panel) suggests. The TAMU model, which in pp corresponds to the SH model + RQM of Fig.~\ref{fig:lcToDzeromodels}, in Pb--Pb collisions implements a coalescence hadronisation process based on the Resonance Recombination Model, includes SMC, and assumes the existence of higher-mass excited states from RQM~\cite{He:2019vgs}. It well reproduces the data. 
\begin{figure}[ht!]
\centering
\includegraphics[width=0.4\textwidth]{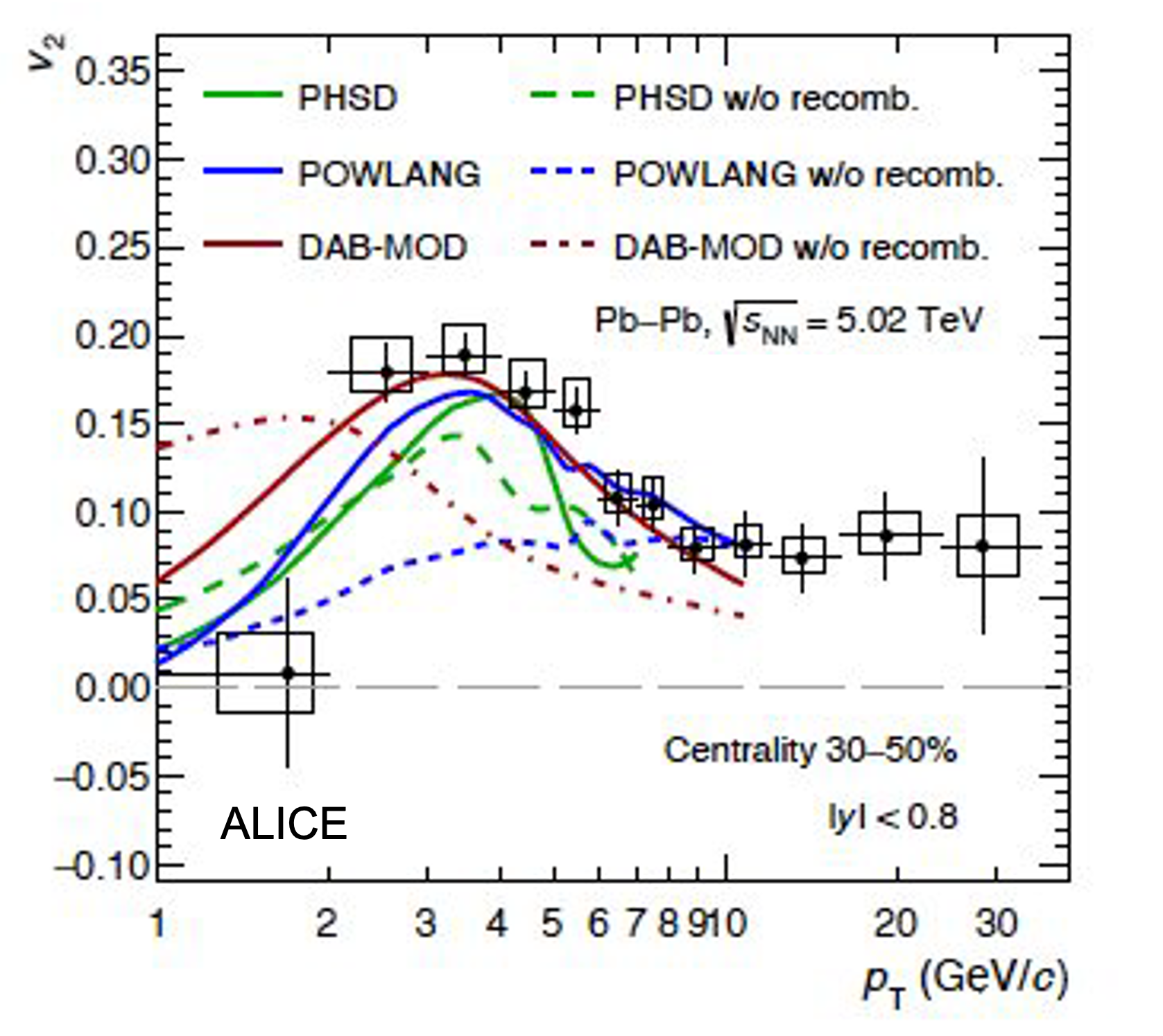}
\includegraphics[width=0.42\textwidth]{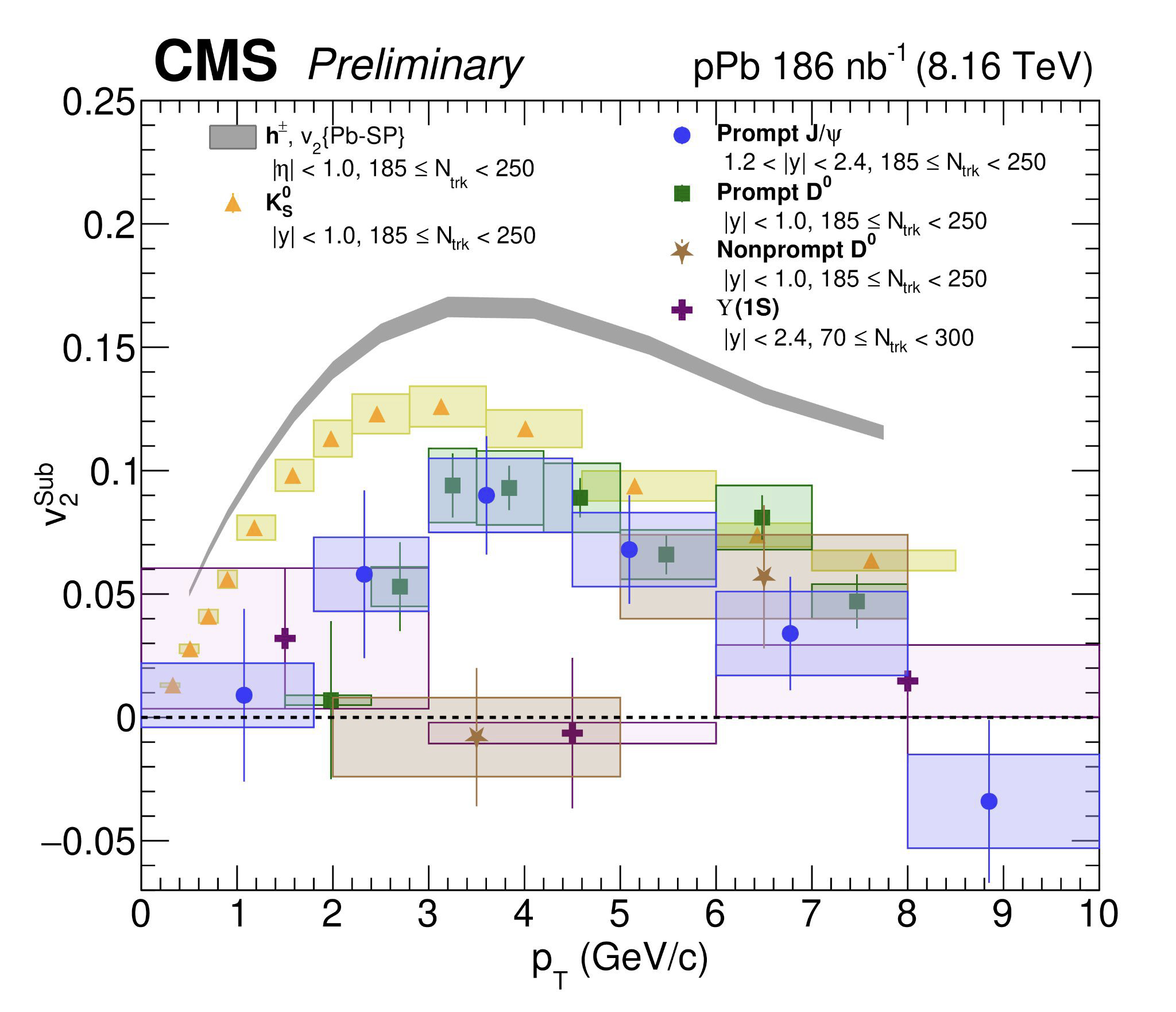}
\caption{Left (from~\cite{ALICE:2021rxa}): comparison of D meson $\vtwo$ with theoretical predictions~\cite{Song:2018tpv,Prado:2016szr,Beraudo:2022dpz} with and without recombination. Right: comparison of prompt and non-prompt \Dzero, prompt $\Jpsi$, \UpsilonOneS, charged-hadron, and $\mathrm{K^{0}_{s}}$ $\vtwo$ measured by CMS in p--Pb collisions~\cite{KLeeHPProceedings}.} 
\label{fig:DmesonFlow}
\end{figure}
The Catania~\cite{Plumari:2017ntm} and POWLANG-HTL~\cite{Beraudo:2022dpz} models qualitatively reproduce the trend though underestimating the data. The GSI/Hd+BW model~\cite{Andronic:2021erx}, based on SH for determining particle abundances and on core-corona with a blast-wave \pt spectrum for the core and a fragmentation-based one for the corona, underestimates the data at intermediate \pt if the RQM higher-mass states are not included. The details of the hadronisation process are not only important for determining the baryon-to-meson ratio but are also fundamental to describe the \Raa and elliptic flow (\vtwo) of D mesons. As an example, in Fig.~\ref{fig:DmesonFlow} (left panel), the importance of including recombination to reproduce D-meson \vtwo data is shown for the POWLANG, PHSD~\cite{Song:2015ykw}, and DAB-MOD~\cite{Prado:2016szr} models. The impact of SMC is discussed extensively in Refs.~\cite{He:2019vgs,Beraudo:2023fpq,CMS:2020bnz}. The impact of recombination and SMC in small systems, where a substantial \vtwo was observed for prompt $\Dzero$ and prompt $\Jpsi$ (right panel of Fig.~\ref{fig:DmesonFlow}, has still to be quantified. Future measurements of charm-baryon \vtwo in p--Pb and Pb--Pb collisions could be very important in this regard.  

The ratio of strange to non-strange meson yields is modified in nuclear collisions. The $\Ds/\Dzero$ measured by ALICE in central Pb--Pb at $\snn=5.02$~\TeV, shown in Fig.~\ref{fig:LcPbPbAllexpModels} (right panel), is higher than that measured in pp in $4<\pt<8$~\GeVc but smaller (at all \pt) than the values reported by STAR, close to 0.4, in $1.5<\pt<5$~\GeVc in semicentral and peripheral Au--Au collisions at $\snn=200$~\GeV~\cite{STAR:2021tte}. The Catania model reproduces ALICE data, which are overestimated by TAMU and POWLANG~HTL, while GSI/Hd+BW does not catch the \pt trend. In the beauty sector, hints that the ratio of strange-meson to non-strange-meson production could be enhanced in Pb--Pb collisions were obtained by CMS~\cite{CMS:2021mzx}, which measured the $\Bs/\Bzero$ ratio, and by ALICE, indirectly, via the non-prompt $\Ds/\Dzero$ ratio~\cite{ALICE:2022xrg}. New measurements with reduced uncertainty and lower \pt reach, as well as the measurement of \XicPlusZero production (larger than \Ds one in pp collisions), are necessary for a final assessment on the enhancement of heavy-flavour strange-particle production. 
\section{Outlook}
The main open points and the most promising next experimental steps, in the author's opinion, to be addressed for understanding heavy-flavour hadronisation were indicated in the previous sections.  
They certainly do not represent a comprehensive list, which is hard to fill considering the many opportunities that the recent experiment upgrades offer, also for quarkonia and exotic states not discussed in these proceedings (see also~\cite{ADublaHPProceedings}). Furthermore, future projects like ALICE3 may open radically new possibilities giving access to multi-charm baryons~\cite{AGrelliHPProceedings}.


\bibliographystyle{utphys}   
\bibliography{bibliography}

\end{document}